\documentclass[pdflatex,sn-mathphys-num]{sn-jnl}

\usepackage{graphicx}
\usepackage{multirow}
\usepackage{amsmath,amssymb,amsfonts}
\usepackage{amsthm}
\usepackage{mathrsfs}
\usepackage[title]{appendix}
\usepackage{xcolor}
\usepackage{textcomp}
\usepackage{manyfoot}
\usepackage{booktabs}
\usepackage{algorithm}
\usepackage{algorithmicx}
\usepackage{algpseudocode}
\usepackage{listings}

\usepackage{comment}
\usepackage{booktabs}
\usepackage{balance}
\usepackage{stmaryrd}
\usepackage{pifont}
\usepackage{listings}
\usepackage{array}
\usepackage{float}
\usepackage{epsfig}
\usepackage{multirow}
\usepackage{url}

\usepackage{multirow}
\usepackage{natbib}
\usepackage{graphicx}
\usepackage{fancyvrb}
\usepackage{subcaption}

\usepackage{listings}
\usepackage{framed}
\usepackage{xcolor}
\usepackage{color}
\colorlet{shadecolor}{gray!20}

\definecolor{shadecolor}{RGB}{220,220,220}

\definecolor{inputcolor}{RGB}{255,139,35}
\definecolor{outputcolor}{RGB}{120,212,252}
\definecolor{embedcolor}{RGB}{254,127,156}
\definecolor{maskcolor}{RGB}{122,128,255}
\definecolor{ecolor}{RGB}{58,149,54}

\definecolor{highcolor}{RGB}{255,153,153}
\definecolor{midcolor}{RGB}{255,204,204}
\definecolor{lowcolor}{RGB}{204,229,255}

\usepackage{tikz}
\usetikzlibrary{shapes,snakes}
\usetikzlibrary{calc}

\usepackage[export]{adjustbox}

\definecolor{green}{RGB}{0,128,0}

\definecolor{yellow}{RGB}{255,200,18}

\newcommand{\bi}{\begin{itemize}}
\newcommand{\ei}{\end{itemize}}

\newcommand{\be}{\begin{enumerate}}
\newcommand{\ee}{\end{enumerate}}
\newcommand{\beqn}{\begin{eqnarray*}}
\newcommand{\eeqn}{\end{eqnarray*}}

\makeatletter
    \newcommand\figcaption{\def\@captype{figure}\caption}
    \newcommand\tabcaption{\def\@captype{table}\caption}
\makeatother

\tikzstyle{mybox} = [draw=black, fill=black!5, thick,
   rectangle, rounded corners, inner sep=0pt, inner ysep=6pt]
\tikzstyle{fancytitle} =[fill=black, text=white]
\usepackage{CJKutf8}
\usepackage{graphicx}

\newcommand{\zh}[1]{\begin{CJK}{UTF8}{gbsn}#1\end{CJK}}

\newcommand{\systemName}{CipherOBS}

\theoremstyle{thmstyleone}

\theoremstyle{thmstyletwo}

\theoremstyle{thmstylethree}

\begin{document}
\title[Article Title]{Decoding Ancient Oracle Bone Script via Generative Dictionary Retrieval}

\author{\fnm{Yin} \sur{Wu}}\email{ywu450@connect.hkust-gz.edu.cn}

\author{\fnm{Gangjian} \sur{Zhang}}\email{gzhang292@connect.hkust-gz.edu.cn}
\author{\fnm{Jiayu} \sur{Chen}}\email{jchen161@connect.hkust-gz.edu.cn}
\author{\fnm{Chang} \sur{Xu}}\email{cxu475@connect.hkust-gz.edu.cn}

\author{\fnm{Yuyu} \sur{Luo}}\email{yuyuluo@hkust-gz.edu.cn}

\author{\fnm{Nan} \sur{Tang}}\email{nantang@hkust-gz.edu.cn}

\author*{\fnm{Hui} \sur{Xiong}}\email{xionghui@hkust-gz.edu.cn}

\affil{\orgdiv{Information Hub}, \orgname{The Hong Kong University of Science and Technology (Guangzhou)}, \state{Guangzhou}, \country{China}}

\abstract{
Understanding humanity's earliest writing systems is crucial for reconstructing civilization's origins, yet many ancient scripts remain undeciphered. Oracle Bone Script (OBS) from China's Shang dynasty exemplifies this challenge: only approximately 1,500 of roughly 4,600 characters have been decoded, and a substantial portion of these 3,000-year-old inscriptions remains only partially understood. Limited by extreme data scarcity, existing computational methods achieve under 3\% accuracy on unseen characters---the core palaeographic challenge. We overcome this by reframing decipherment from classification to dictionary-based retrieval. Using deep learning guided by character evolution principles, we generate a comprehensive synthetic dictionary of plausible OBS variants for modern Chinese characters. Scholars query unknown inscriptions to retrieve visually similar candidates with transparent evidence, replacing algorithmic black boxes with interpretable hypotheses. Our approach achieves 54.3\% Top-10 and 86.6\% Top-50 accuracy for unseen characters. This scalable, transparent framework accelerates decipherment of a pivotal undeciphered script and establishes a generalizable methodology for AI-assisted archaeological discovery.

}


\maketitle

\section{Main}
\label{sec:intro}
The decipherment of ancient scripts offers a direct window into the origins of human civilization~\citep{assael2022restoring, marchant2025ai, assael2025contextualising}. Oracle Bone Script (OBS), which emerged more than 3,000 years ago during the latter half of the Shang dynasty (c.\,1250--1046 BCE), is particularly significant among these ancient writing systems. Inscribed on turtle plastrons and animal bones for divination purposes, these glyphs represent the earliest known form of Chinese writing. They provide detailed records of early society that range from rituals and warfare to astronomical observations and governance. Archaeologists have excavated approximately 160,000 fragments bearing roughly 4,600 distinct characters from this foundational period of Chinese history~\citep{bazerman2009handbook, wang2024open}. Despite more than a century of research, however, only approximately 1,500 characters have been deciphered~\citep{wang2024open}. A substantial portion of the script thus remains unread, limiting our understanding of early societal evolution and the transition of humanity into recorded history~\citep{boltz1986early, keightley1979shang, li2025comprehensive}.

Palaeography, the study of ancient writing, confronts significant challenges in the decipherment of OBS. Characters are frequently eroded, fragmented, or exhibit considerable stylistic variation---an inherent feature of pre-modern writing compounded by the physical effects of incision on bone and shell and three millennia of material degradation. Furthermore, graphic evolution over this span obscures the connections between these ancient forms and modern Chinese script~\citep{zhen1995astronomy, takashima2000towards, chen2009compound}. One important palaeographic strategy involves identifying structural parallels by comparing strokes, components, and evolutionary patterns to hypothesize modern equivalents. This method situates undeciphered glyphs within a network of historical and structural relationships, grounding interpretations in empirical evidence rather than speculation. However, structural comparison is far from the only strategy employed: phonological, contextual, and philological analyses all play crucial roles in decipherment~\citep{takashima2000towards}. Moreover, many OBS graphs are not structurally isomorphic to any modern character, because the words they wrote became obsolete or because the character forms diverged substantially over time. The manual process is exceptionally laborious and demands specialized expertise for comparisons against extensive dictionaries and archaeological records. As a result, progress has been slow, limiting the investigation of this foundational civilization.

Machine learning is beginning to transform this field by enabling large-scale digitization~\citep{guo2015building, liu2020oracle}, character recognition~\citep{li2011recognition, meng2017recognition, gan2023characters}, and initial attempts at automated OBS decipherment~\citep{guan2024deciphering}. Nevertheless, computational approaches face a fundamental mismatch between data availability and task demands. The entire corpus of approximately 160,000 excavated fragments bearing roughly 4,600 distinct characters must be matched against the full set of Unicode-encoded CJK ideographs, which exceeds 87,000 entries~\citep{zhang2025megahan97k}---a set that encompasses characters from all historical periods and regional traditions, far exceeding the few thousand characters in active modern use. This discrepancy creates an extreme imbalance: for every ancient character in the dataset, the model must distinguish among numerous potential modern equivalents, most of which lack attested ancient predecessors. Thousands of entries in this search space exist only as blank labels, which forces classification models to make distinctions without evidence---a fundamentally intractable challenge for supervised approaches.

This data imbalance leads directly to failed generalization, which is precisely the scenario archaeologists encounter when discovering new fragments. State-of-the-art classification methods such as OBSD~\citep{guan2024deciphering} achieve less than 3\% Top-1 accuracy on previously unseen OBS characters in our standard benchmarks. The root cause is structural. Without training examples of a novel glyph, supervised models cannot extrapolate plausible modern equivalents. This stands in sharp contrast to handwriting recognition, where fixed character categories and abundant per-class examples enable reliable classification. OBS decipherment operates in an open-world setting where the central task is interpreting characters outside the training vocabulary. These characters, by definition, lack labels for learning. Consequently, the classification paradigm is designed for closed and data-rich scenarios, making it fundamentally misaligned with the discovery-driven nature of archaeological research.

We address these limitations by reframing the problem from classification to dictionary-based retrieval. Rather than forcing models to predict a single label from tens of thousands of candidates, we generate a comprehensive synthetic dictionary containing plausible OBS-style variants of modern Chinese characters. We then search this dictionary for visual matches. When the system queries an unknown inscription, it returns candidate correspondences alongside generated OBS variants that explain how a modern character could yield the observed ancient form. This approach offers two critical advantages. First, it transforms an opaque algorithmic decision into interpretable and evidence-based hypotheses that scholars can validate, thereby addressing the black-box concern inherent in deep learning. Second, retrieval is inherently robust to the noise, damage, and variation typical of archaeological artifacts because matching relies on structural similarity rather than pixel-perfect classification. By converting limited data into a generative format, we provide a scalable pathway for decipherment.

\section{Results}

\subsection{Decoding Oracle Bone Script}

\begin{figure}[t]
    \centering
    \includegraphics[width=\linewidth]{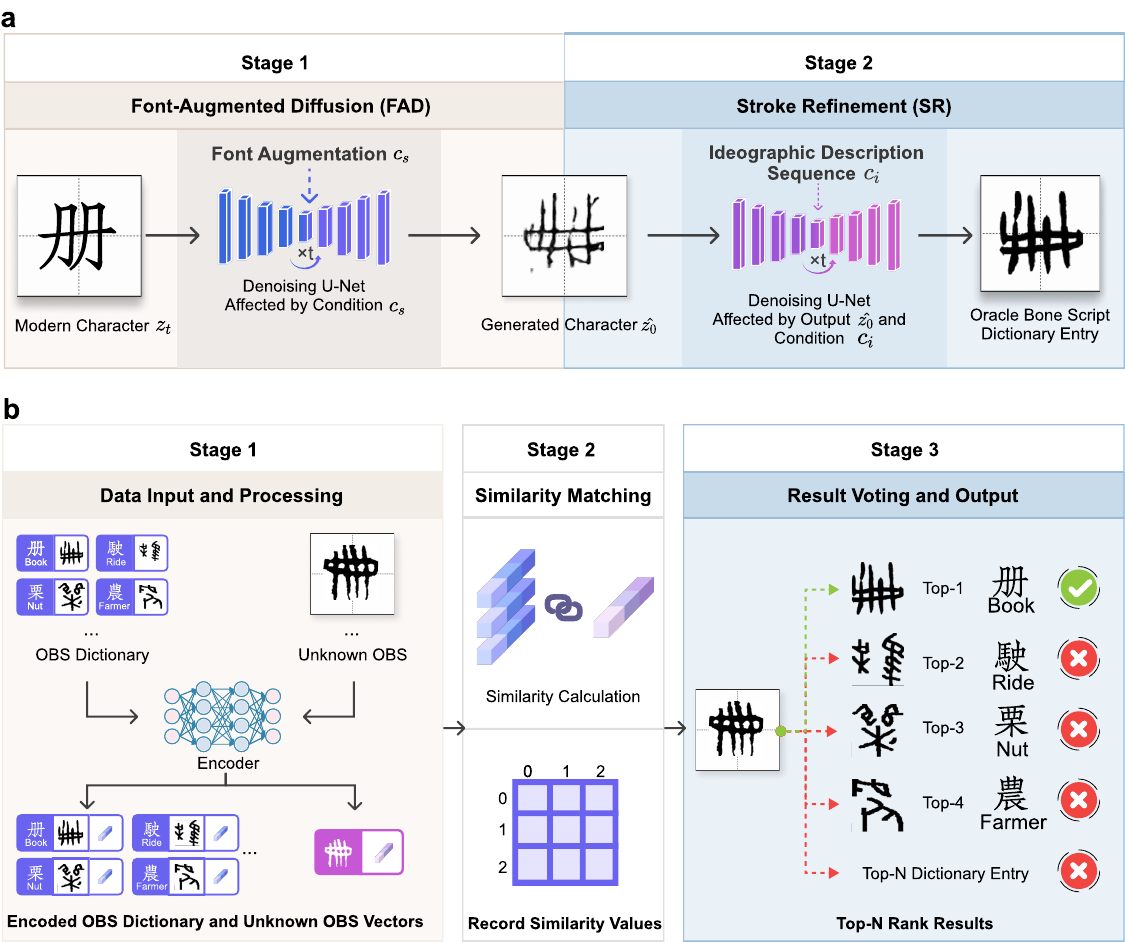}
    \caption{Overview of the {\systemName} system for synthesizing and retrieving OBS variants. \textbf{a,} Two-stage synthesis process: FAD generates initial OBS drafts from modern Chinese character images using a denoising U-Net conditioned on font-augmented inputs; SR then refines these drafts with IDS, derived from the modern character to produce high-fidelity variants, accounting for historical erosion and stylistic differences. \textbf{b,} Dictionary-based retrieval: A encoder embeds the input OBS image and synthesized dictionary entries into a feature space; cosine similarity identifies Top-$k$ matches, with voting on modern character labels for robustness against degradation. This pipeline creates a visual dictionary of OBS-modern pairs for scalable decipherment. English translations are also provided for each OBS's Chinese explanations.}
    \label{fig:evaluation_process}
\end{figure}

To bridge the decipherment gap, we developed {\systemName}, a framework that reframes OBS decoding as a retrieval task powered by a generative dictionary. Analogous to a cybersecurity ``dictionary attack,'' our system synthesizes a comprehensive library of plausible OBS variants for modern Chinese characters by learning evolutionary patterns via diffusion. When presented with an undeciphered glyph, {\systemName} queries this synthetic corpus to retrieve ranked equivalents. This approach achieves state-of-the-art performance, generalizes to unseen characters, and proves robust to the degradation that confounds standard classifiers.

An interactive pipeline supports scholarly review by presenting visual hypotheses and evolutionary cues. As a comprehensive, artificial intelligence (AI)-generated OBS dictionary designed for human interpretability, {\systemName} enables collaborative, scalable analysis in paleography and accelerates insights into early civilizations.

\subsubsection{A comprehensive generative script dictionary}

The dictionary underlying {\systemName} is synthesized from modern Chinese characters to compensate for the scarcity and uneven quality of excavated OBS exemplars.
The design is deliberately two‑stage (Fig.~\ref{fig:evaluation_process}a), reflecting properties of Chinese writing and OBS paleography documented in the literature: characters are compositional~\citep{wang1999reading, reichle2018models, Boltz+2021+845+864}, built from recurrent components arranged by stable spatial relations, and these structural regularities persist across historical forms~\citep{boltz1986early, takashima2000towards}.
Meanwhile, OBS glyphs were incised on bone or shell and often exhibit erosion, stroke attrition, and carving irregularities that introduce substantial stylistic variability. Decoupling global appearance from structural fidelity, therefore, yields a principled synthesis strategy: the first stage transfers overall morphology into an OBS‑like style, and the second stage enforces component‑wise plausibility and layout consistency using explicit structural guidance.

Stage~1, Font‑Augmented Diffusion (FAD), applies a conditional diffusion model~\citep{ho2020denoising} with a denoising U‑Net backbone~\citep{ronneberger2015u}. Modern characters are rendered across diverse fonts to expose the model to shape variability while preserving the recurrent component structure characteristic of Chinese script. Conditioned on these renderings, FAD produces an OBS‑style draft that captures the representative outlines and efficient stroke minimization of incised inscriptions without requiring large labeled ancient corpora.

Stage~2, Stroke Refinement (SR), sharpens and regularizes the draft with guidance from the character’s Ideographic Description Sequence (IDS), using an IDS‑guided diffusion refinement adapted from recent font‑editing architectures~\citep{yang2024fontdiffuser}. IDS represents a character as a composition of components linked by spatial operators (for example, left-right, top-bottom, enclosure). Conditioning on this blueprint constrains refinement to preserve component identity and layout while permitting historically attested stroke‑level changes such as attrition, merging, and curvature variation. The result is a small set of variants for each modern character that reflect stylistic diversity and inscriptional wear observed in OBS corpora. Each dictionary entry stores the synthetic OBS image together with its modern label, enabling visual and structural auditing during retrieval.

\subsubsection{Dictionary-based retrieval of ancient scripts}

With the dictionary in place, {\systemName} retrieves matches for input OBS images via Dictionary-based OBS Retrieval. In retrieval stage~1 (data input and processing), a shared ConvNeXt encoder~\citep{liu2022convnet} embeds both the query OBS image and every dictionary entry into a common feature space, emphasizing ideographic structure while reducing sensitivity to erosion and carving style.
Stage~2 (similarity matching) in retrieval computes cosine similarity between the query embedding and all dictionary embeddings, yielding a ranked list of candidates.
Stage~3 (result voting and output) in retrieval aggregates labels across the top‑ranked synthetic variants using a simple voting rule to mitigate intra‑class variation and partial degradation. The system returns the Top-$N$ results together with exemplar OBS images, allowing scholars to verify structural correspondences---such as component layout and stroke economy---rather than relying solely on similarity scores. This pairing of visual evidence with an explicit structural rationale supports expert scrutiny and reproducible, audit‑ready hypotheses about undeciphered glyphs.

\subsection{Benchmarking {\systemName} decipherment performance}

\subsubsection{Quantitative assessment of {\systemName} accuracy}

We evaluate the proposed {\systemName} on a composite benchmark derived from HUST-OBS~\citep{wang2024open} and EVOBC~\citep{guan2024open}, the two largest public OBS resources. The benchmark contains 71{,}698 high-fidelity rubbings covering 1{,}590 deciphered OBS characters, each mapped to a modern Chinese character equivalent. We enforce a zero-shot split at the character level: 90\% of characters are used for training and 10\% for testing, with no label overlap between splits, mirroring scenarios where newly excavated glyphs correspond to unseen characters.

Since no prior method directly corresponds to our dictionary-based retrieval approach, our baselines follow the ``Generate Modern Characters from OBS---OCR'' paradigm introduced by~\cite{guan2024deciphering}. These include image-generation models such as CycleGAN \citep{zhu2017unpaired}, diffusion-based variants (e.g., CDE \citep{saharia2022image}, Conditional Diffusion \citep{dhariwal2021diffusion}, BBDM \citep{li2023bbdm} and OBS Decipher itself (OBSD; \citep{guan2024deciphering})), and a direct retrieval (DR) baseline. All methods use identical preprocessing, the same unified 1,590 label space, and the same zero-shot split; training settings for the baselines follow~\cite{guan2024deciphering} to ensure fairness. All generated images were classified using the OBS optical character recognition (OCR) tool from \citep{guan2024deciphering}, reported to achieve 99.87\% accuracy on 88,899 Chinese character categories. Each method outputs a ranked list over the shared label space.
We report Top-$N$ accuracy ($N = 1, 10, 20, 50, 100$) as the fraction of test samples whose correct label appears within the Top-$N$ predictions. Our approach performs dictionary-based retrieval over a synthetic OBS variant dictionary generated from modern characters.
We report two variants of our method: a single-pass dictionary and a 20-iteration refinement. In each iteration, stroke-level refinement updates synthesized variants and their embeddings while preserving correct dictionary entries; test inputs remain unchanged, and rankings are recomputed on the updated dictionary.

\renewcommand{\arraystretch}{1.5}
\begin{sidewaystable}
\caption{Performance comparison of OBS decipherment methods across classification, retrieval and dictionary-based retrieval paradigms on HUST-OBS and EVOBC.}\label{tab:performance_comparison}
\begin{tabular*}{\textheight}{@{\extracolsep\fill}lcccccccccc}
\toprule
& \multicolumn{5}{@{}c@{}}{\textbf{Generate Modern Characters from OBS---OCR}} & \multicolumn{1}{@{}c@{}}{\textbf{Retrieval}} & \multicolumn{2}{@{}c@{}}{\textbf{Dictionary-based Retrieval}} \\
\cmidrule(lr){2-6} \cmidrule(lr){7-7} \cmidrule(lr){8-9}
\textbf{Rank} & \textbf{CDE} & \textbf{CycleGAN} & \textbf{BBDM} & \textbf{Conditional Diffusion} & \textbf{OBSD} & \textbf{DR} & \textbf{\systemName} & \textbf{\systemName} (20 iterations) \\
\midrule
Top-1 acc.    & 0.0\%  & 0.0\%  & 0.0\% & 2.5\% & 1.9\% & 7.9\%  & 21.2\% & 47.5\% \\
Top-10 acc.   & 0.0\%  & 0.1\%  & 0.0\% & 3.0\% & 2.8\% & 15.1\% & 54.3\% & 79.6\% \\
Top-20 acc.   & 0.0\%  & 0.2\%  & 0.1\% & 3.1\% & 3.0\% & 19.6\% & 66.8\% & 89.3\% \\
Top-50 acc.   & 0.0\%  & 0.3\%  & 0.1\% & 3.2\% & 3.2\% & 27.7\% & 86.6\% & 98.5\% \\
Top-100 acc.  & 3.0\%  & 0.5\%  & 0.2\% & 3.2\% & 3.4\% & 36.1\% & 96.9\% & 99.9\% \\

\bottomrule
\end{tabular*}
\footnotetext{Performance comparison on the character split (1{,}590 labels; 10\% test characters; no label overlap). All methods share identical preprocessing, the same split and a unified OCR tool. Baselines follow the OBSD ``Generate Modern Characters from OBS---OCR'' paradigm (CDE, CycleGAN, Conditional Diffusion, BBDM, OBSD); DR is a direct retrieval baseline; Our method is dictionary-based retrieval. The 20-iteration variant applies iterative dictionary refinement while preserving correct entries. Top-$N$ accuracy denotes the percentage of test samples whose correct label appears within the Top-$N$ ranked predictions.
}
\end{sidewaystable}

The results reveal a consistent divide between classification-style pipelines and retrieval. The image-to-image generation model (CycleGAN) and diffusion variants trained under the traditional paradigm underperform in the zero-shot setting. This pattern reflects the difficulty of generalizing from seen to unseen characters: errors in the generative translation step propagate to OCR, and domain shift between training labels and unseen test labels limits extrapolation~\citep{guan2024deciphering}. Additional discussion of baseline behavior is provided in Section~\ref{sec:disc}.

By contrast, {\systemName} exhibits a markedly different profile. It attains 21.2\% Top-1 and scales strongly with $N$: 54.3\% (Top-10), 86.6\% (Top-50), and 96.9\% (Top-100)---a 75.7 percentage points (pp) gain from Top-1 to Top-100. Iterative dictionary refinement further improves Top-100 accuracy to 99.9\% after 20 iterations. Relative to OBSD, the absolute improvement at Top-10 is 51.5 pp. Compared with DR, which reaches $27.7\%$ Top-50 and $36.1\%$ Top-100, {\systemName} benefits from an overcomplete dictionary (multiple plausible variants per character), yielding substantially higher recall at practical cutoffs.

Conceptually, retrieval over a synthesized dictionary shifts the burden from learning decision boundaries for unseen labels to matching against an enumerated, interpretable hypothesis set of plausible OBS-style variants. Practically, high Top-50/Top-100 accuracy reduces an expert’s search space from 1{,}590 labels to 50--100 candidates (a reduction by a factor of approximately 31.8 to 15.9) while preserving interpretability through visual correspondences. This combination of accuracy and transparency supports palaeographic workflows and accelerates hypothesis formation on undeciphered characters. The contribution of each component is analyzed in ablation studies.

\subsubsection{Synthesizing authentic oracle bone script variants}
\label{ssec: authentic obs}

\begin{figure}[t]
    \centering
    \includegraphics[width=\linewidth]{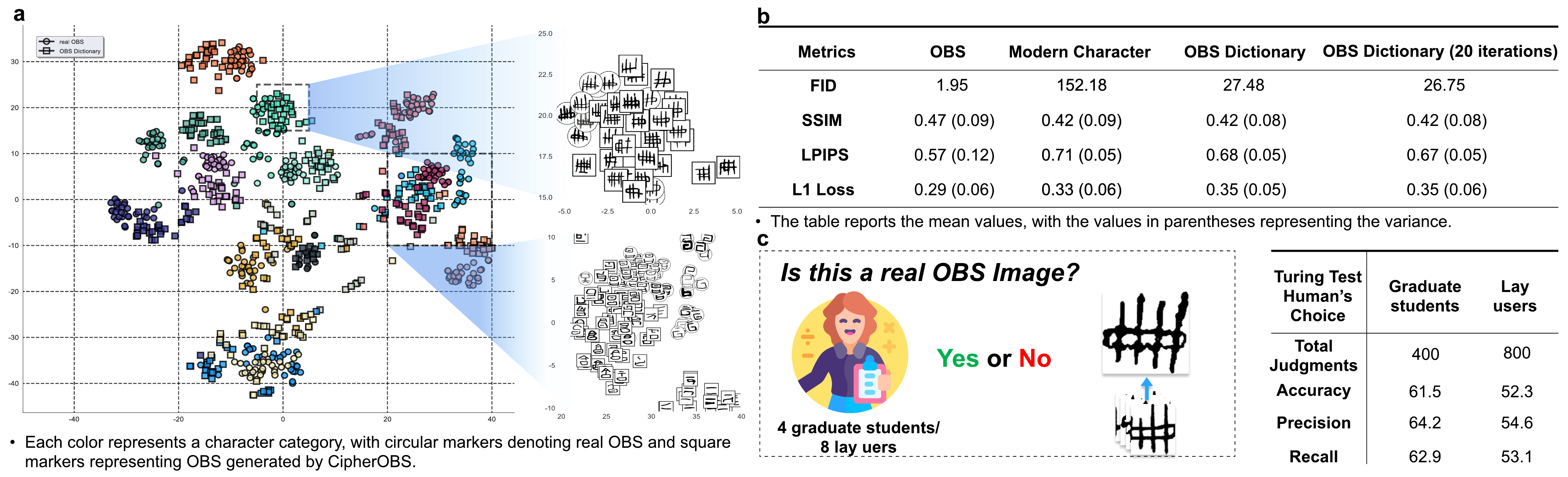}
    \caption{Validation of the visual fidelity of the synthetic OBS dictionary.
    \textbf{a,} 2D t‑SNE of retrieval‑encoder features for 15 characters. Authentic glyphs (circles) and synthetic counterparts (squares) co‑localize within character‑specific clusters; insets enlarge two regions.
    \textbf{b,} Quantitative comparison between real OBS images and three candidate sets: modern character renderings, entries from our OBS dictionary, and dictionary entries after 20 refinement iterations. Metrics: FID, LPIPS, L1 loss (lower is better); SSIM (higher is better); Values are means across 20 runs; standard deviations are in parentheses.
    \textbf{c,} Results of a Turing‑style discrimination task (n = 12 participants; 1{,}200 judgments; balanced 5 real/5 synthetic per set).}
    \label{fig:dictionary_verify}
\end{figure}
For retrieval to be reliable, synthesized dictionary entries must approximate the visual and structural properties of OBS. We therefore evaluate the synthetic dictionary using three complementary analyses---feature‑space visualization, image‑level similarity metrics, and human discrimination---applied to the dictionary outputs (Fig.~\ref{fig:dictionary_verify}).

We first examine the feature space distribution of real and synthetic glyphs. We randomly select 15 OBS character types, each with authentic examples and dictionary‑generated variants, and embed them with 2D t‑distributed stochastic neighbor embedding (t‑SNE) with perplexity 30 and a cosine distance metric using features from the retrieval ConvNeXt encoder. As shown in Fig.~\ref{fig:dictionary_verify}a, real (circles) and synthetic (squares) samples for a given character co‑localize into mixed clusters. Because t‑SNE is qualitative and does not preserve global geometry, we interpret this co-localization as suggestive alignment of real and synthetic exemplars in the retrieval feature space rather than as a statistical test. Occasional mixed clusters among historically related characters are consistent with shared components and known palaeographic relationships, and do not necessarily indicate synthesis errors.

Next, we quantify fidelity using four established metrics that span perceptual and pixel-level similarity~\citep{yang2024fontdiffuser, zhu2017unpaired, heusel2017gans}: Fréchet Inception Distance (FID), Learned Perceptual Image Patch Similarity (LPIPS), Structural Similarity Index Measure (SSIM) and L1 loss, where lower values are better for FID, LPIPS and L1 loss, and higher values are better for SSIM). Figure~\ref{fig:dictionary_verify}b reports means across 20 runs (dispersion in parentheses). A self‑consistency baseline comparing splits of real OBS images yields FID 1.95, which calibrates the attainable lower bound in this domain. Modern character renderings are far from the OBS distribution (FID 152.18). Our OBS dictionary substantially narrows this gap (FID 27.48), with a modest additional improvement after 20 refinement iterations (FID 26.75). Perceptual trends are consistent in LPIPS (modern 0.71; dictionary 0.68; after 20 iterations 0.67). Pixel‑level metrics show smaller differences: SSIM is comparable across modern and dictionary images (0.42), and L1 is slightly higher for the dictionary (0.35 vs. 0.33 for modern). This pattern is expected: pixel‑aligned measures penalize legitimate structural and compositional shifts between eras, whereas distributional and perceptual metrics are better aligned with stylistic fidelity and texture.

Finally, we assess perceptual realism with a Turing‑style discrimination task (12 participants: 8 lay users, 4 OBS‑trained graduate students; 1{,}200 total judgments; Fig.~\ref{fig:dictionary_verify}c). Each trial presents 10 images (5 authentic, 5 synthetic), and participants label each image as real or generated (chance accuracy = 50\%). Lay participants perform near chance (accuracy 52.3\%, precision 54.6\%, recall 53.1\%), while graduate students are modestly above chance (accuracy 61.5\%, precision 64.2\%, recall 62.9\%). These outcomes indicate that non-experts find synthetic variants difficult to distinguish from authentic inscriptions, with experts detecting subtle artifacts. We note that the lay user results should be interpreted with caution: participants without OBS exposure may lack the visual criteria needed to discriminate between real and poorly generated variants, so near-chance performance does not by itself confirm high fidelity. The graduate student results, showing modestly above-chance detection, provide a more meaningful assessment.

Taken together, the qualitative co‑localization in retrieval features, the favorable perceptual metrics, and the human assessments indicate that the synthetic dictionary captures salient properties of OBS and is suitable as a proxy for retrieval.

\subsection{Evaluating {\systemName} in the real world}

\subsubsection{Generalizing to newly deciphered glyphs}

\begin{figure}[t]
\centering
\includegraphics[width=\linewidth]{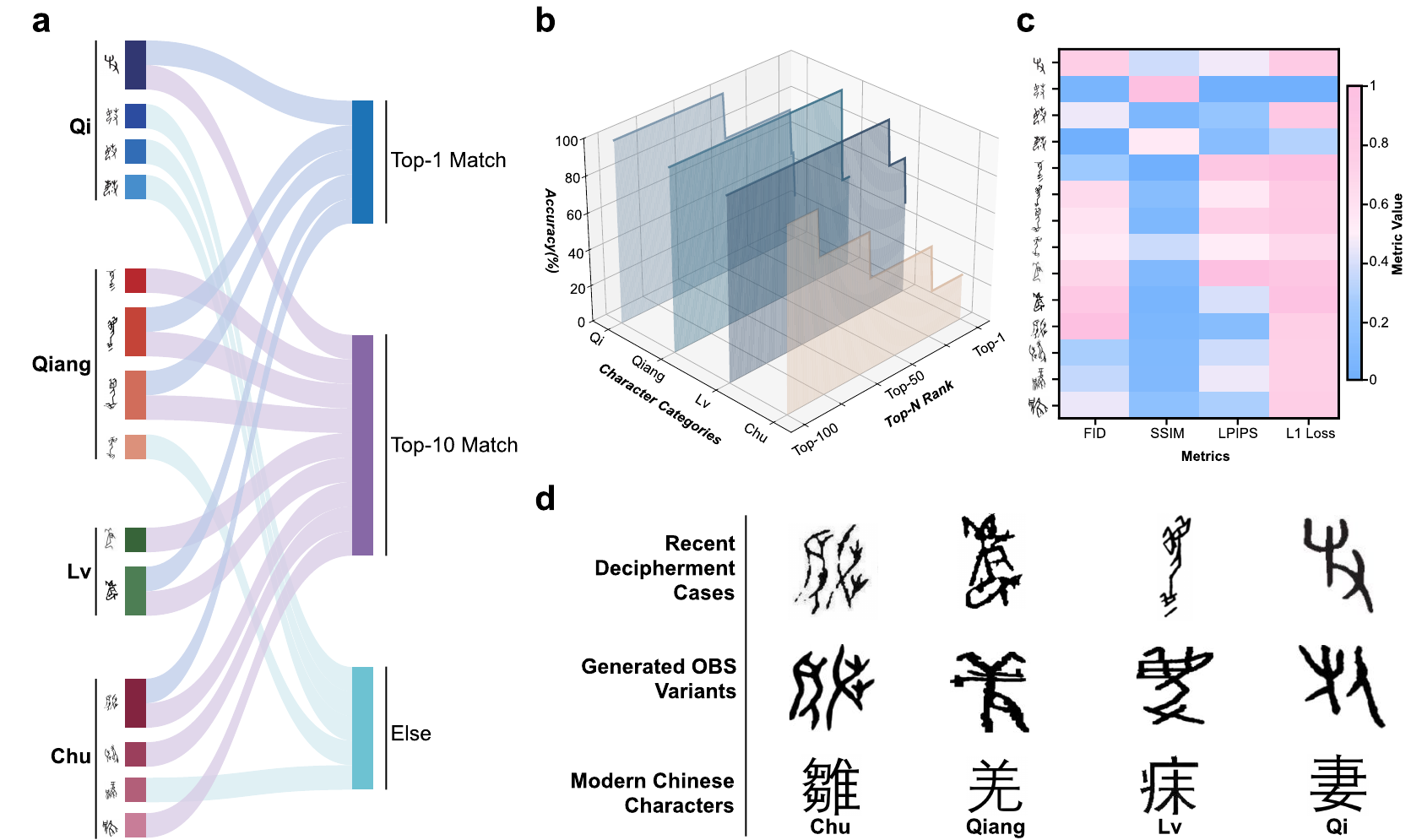}
\caption{Zero-shot generalization to newly deciphered script characters.
\textbf{a,} Sankey diagram summarizing retrieval outcomes for 14 glyphs grouped by graph group---Qi (`wife'; \zh{妻}), Qiang (`Qiang people'; \zh{羌}), L\"{u} (`shoe'; \zh{履}), Chu (`chick'; \zh{雛}). Flows indicate whether the structurally matched modern character appears at Top-1, within the Top-10, or outside the Top-10 shortlist.
\textbf{b}, Top-$N$ structural matching accuracy as a function of $N$; accuracy reaches 100\% within $N=100$, consistent with expert-in-the-loop inspection of ranked candidates.
\textbf{c,} Quality metrics for the corresponding synthesized dictionary entries, min-max normalized across the 14 cases. For FID, LPIPS, and L1 loss, lower values indicate better alignment; for SSIM, higher is better. Perceptual similarity remains comparatively strong across categories, while pixel-aligned agreement is weaker for Chu, reflecting erosion and stylistic variability.
\textbf{d,} Qualitative links between recently studied OBS glyphs (top), synthesized OBS-style variants (middle), and modern Chinese characters (bottom) for Chu, Qiang, L\"{u}, and Qi. We note that structural similarity to a modern character is a hypothesis-generating step; full decipherment requires broader philological analysis including phonological and contextual evidence.}
\label{fig:newly_deciphered}
\end{figure}

A decisive test for computational decipherment is whether a system assists scholars when confronted with newly excavated, previously unseen inscriptions. We therefore evaluate zero-shot performance on 14 OBS glyphs discussed in post-2022 studies~\citep{Wu2025, NiuLiu2023, Yuan2022, Hou2024} and excluded from all training and dictionary construction. We note that these cited papers represent ongoing scholarly debate; the interpretations they propose are not universally settled decipherments. We use them here to test whether our system retrieves structurally plausible candidates, not to claim definitive decipherment. To ensure a genuinely out-of-distribution setting, all training corpora and the synthetic dictionary were compiled exclusively from sources dated prior to 2022.

Querying the generative dictionary returns a ranked list of candidate modern characters (Fig.~\ref{fig:newly_deciphered}a). Outcomes stratified by graph group---Qi\footnote{Throughout this paper, romanizations follow standard Hanyu Pinyin. ``Lv'' is used as a shorthand for ``L\"{u}'' in contexts where diacritics are unavailable in our labelling system.} (`wife'; \zh{妻}), Qiang (`Qiang people'; \zh{羌}), Lv (`shoe'; \zh{履}), and Chu (`chick/fledgling'; \zh{雛})---show clear trends (Fig.~\ref{fig:newly_deciphered}a). For instance, Qiang exhibits the highest share of Top-1 recoveries, Chu and Lv concentrate within the Top-10, and Qi remains the most challenging, with a greater fraction falling outside the Top-10 shortlist. Top-$N$ accuracy increases and reaches 100\% within $N=100$ (Fig.~\ref{fig:newly_deciphered}b), consistent with expert-in-the-loop workflows that inspect ranked candidates rather than relying on a single prediction.

We further analyze the quality of synthesized dictionary entries associated with these cases using four established metrics (Fig.~\ref{fig:newly_deciphered}c): FID, LPIPS, L1 loss, SSIM.  Under this convention, perceptual dissimilarity metrics (FID, LPIPS) are comparatively favorable across categories, whereas pixel-aligned agreement is weaker in more eroded or stylistically variable cases---especially Chu, which shows lower SSIM and higher L1. Qi presents more balanced signals across perceptual and structural cues, aligning with its superior Top-1 retrieval. These patterns indicate that perceptual alignment, rather than strict pixel-level correspondence, is the principal driver of retrieval under erosion and fragmentary conditions.

Qualitative examples (Fig.~\ref{fig:newly_deciphered}d) link recently studied OBS glyphs (top) to synthesized OBS-style variants (middle) and their modern counterparts (bottom). Retrieved variants exhibit structural similarities---such as shared component shapes or analogous spatial arrangements---that may aid hypothesis formation. We emphasize that structural similarity to a modern character generates a hypothesis to be tested through broader philological analysis; it does not by itself constitute decipherment. Even where pixel-level agreement is weaker (e.g., Chu), retrieved variants may surface structurally plausible candidates for further expert evaluation.

Although based on a modest sample, these patterns mirror broader benchmarking results. Retrieval behavior is governed chiefly by perceptual rather than pixel-aligned similarity---an advantage under erosion and stylistic variability (Fig.~\ref{fig:newly_deciphered}c). By returning ranked, visually interpretable candidates alongside synthesized OBS-style variants, the system supports expert-in-the-loop adjudication (Fig.~\ref{fig:newly_deciphered}d).

\subsubsection{Robustness to archaeological degradation}
\label{sec:degradation}

\begin{figure}[t]
    \centering
    \includegraphics[width=\linewidth]{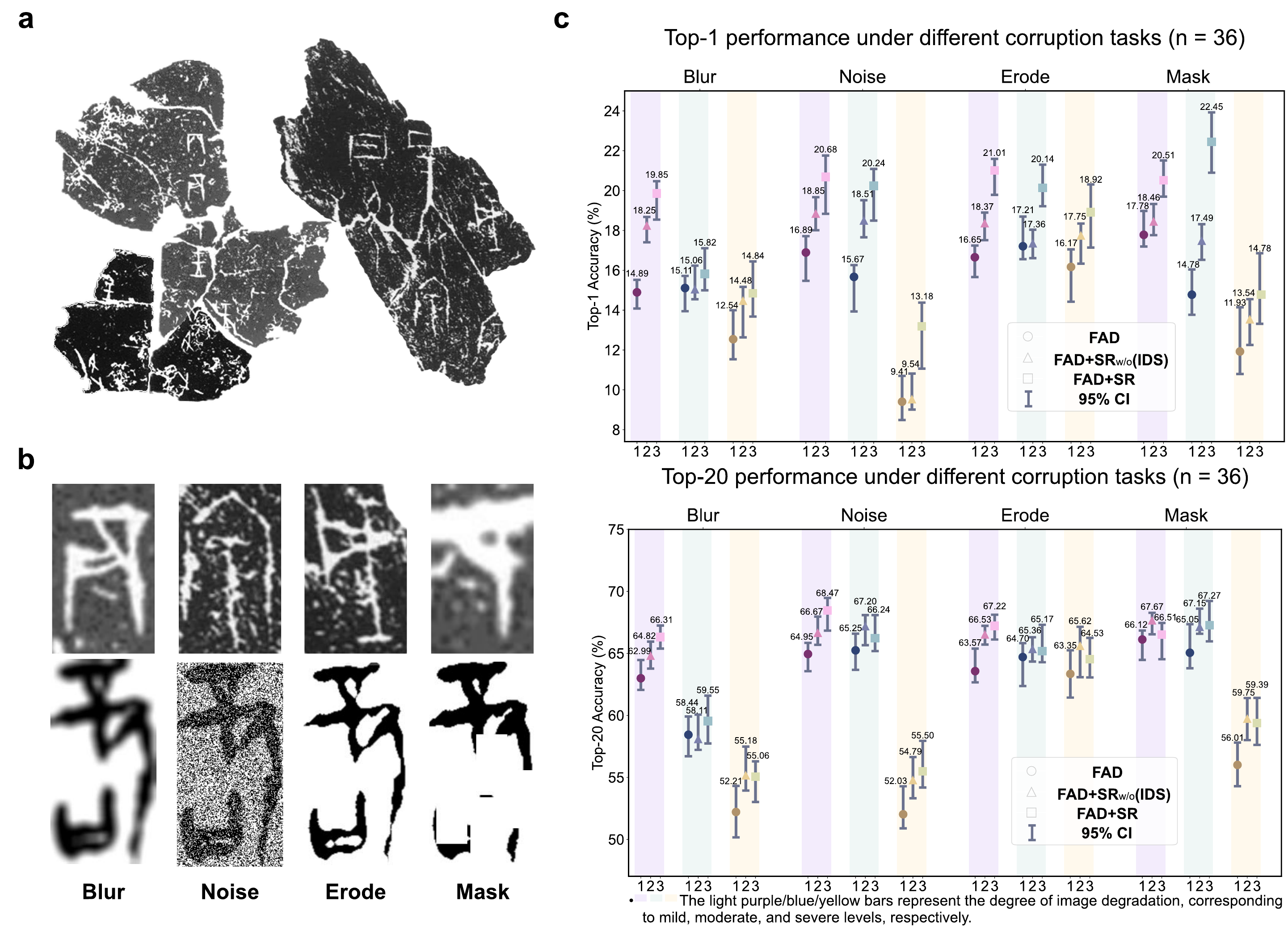}
    \caption{Robustness of {\systemName} to field-realistic image degradation. \textbf{a}, Representative digitized oracle-bone fragments exhibiting common surface conditions in archaeological contexts. \textbf{b}, Degradation types (top: field manifestations; bottom: synthetic operators applied to a test glyph): Blur (Gaussian defocus), Noise (additive Gaussian noise), Erode (stroke thinning/chipping), and Mask (random occlusion), each at three severities. \textbf{c}, Top-1 (upper) and Top-20 (lower) retrieval accuracy for FAD, FAD+SR without IDS and the full system (FAD+SR) across 12 conditions (4 types~$\times$~3 severities). Points indicate mean accuracy across test images; error bars denote 95\% confidence intervals; background shading encodes severity (1\,=\,light; 2\,=\,medium; 3\,=\,heavy).}
    \label{fig:degradation}
\end{figure}

Archaeological OBS fragments frequently exhibit surface and imaging artifacts that complicate decipherment, including erosion of incised strokes, occlusion by neighboring shards, digitization noise, and optical blur (Fig.~\ref{fig:degradation}a), as noted previously~\citep{diao2025oracle, wang2025chinese}. To assess field-realistic performance, we synthetically applied four degradation operators---Blur, Noise, Erode, and Mask---to every test glyph at three severities (light, medium, heavy; Fig.~\ref{fig:degradation}b), using standard image-processing primitives. We compared {\systemName} with two system variants: FAD only; FAD+SR without IDS. Retrieval accuracies are reported at multiple ranks, focusing on Top-1 and Top-20 in Fig.~\ref{fig:degradation}c; on clean scans (no degradation), the full system attains Top-1~21.2\%, Top-10~54.3\%, Top-20~66.8\%, Top-50~86.6\%, and Top-100~96.9\%.

Across all 12 degradation conditions (4 types~$\times$~3 severities), the full system consistently outperformed both ablations for Top-1 and Top-20 accuracy and generally exhibited tighter 95\% confidence intervals (Fig.~\ref{fig:degradation}c). The pattern of robustness aligns with field conditions. Erosion and masking, which are pervasive in excavated material, produced only modest declines relative to clean scans: even at heavy severity, Top-1 decreased moderately while Top-20 remained close to the clean-scan level, indicating that degraded glyphs still retrieve stable candidate sets. By contrast, heavy noise or blur caused the largest drops---Top-1 falling to the low-to-mid teens and Top-20 to the mid-to-high 50\% range---consistent with these operators removing the high-frequency stroke morphology required for precise matching. Mild-to-moderate noise or blur induced only incremental losses, reinforcing that the retrieval process tolerates realistic digitization imperfections. 

The ablations illuminate component roles. Relative to FAD alone, adding SR improved both Top-1 and Top-20 across all degradation types and severities, reflecting the benefit of stroke-aware refinement. Removing IDS guidance consistently reduced performance compared with the full system, with the gap most pronounced under erosion and masking, where maintaining component layout is critical. These trends are consistent with the system design: the synthetic dictionary supplies multiple structurally plausible variants that enable partial matching when strokes thin, chip, or are partially occluded, while SR guided by IDS regularises component topology so that the encoder remains less sensitive to fragmentation and local attrition. Under extreme noise or blur, however, most discriminative morphology is suppressed, revealing a practical legibility threshold for reliable retrieval.

Practically, these results indicate that dictionary-based retrieval preserves the quality of Top-$N$ hypotheses under the degradations most typical of field scans. For erosion and occlusion in particular, the full system maintains near-clean Top-20 accuracy, ensuring that experts receive stable, interpretable candidate lists even when single-best matches are uncertain.

\subsubsection{Enhancing human-AI decipherment}

\renewcommand{\arraystretch}{1.5}
\begin{table}[t]
\centering
\caption{Human evaluation of OBS interpretation with and without retrieval-based assistance. Values are mean~$\pm$~s.d. across participants (n = 12; 50 trials per participant). Response time is per trial in seconds; confidence is self-reported per trial on a 1--5 Likert scale (1 = very uncertain, 5 = very certain). Differences are absolute (pp for accuracy, seconds for time, points for confidence).}
\label{tab:human_evaluation_new}
\setlength{\tabcolsep}{5pt}
\begin{tabular*}{\textwidth}{@{\extracolsep{\fill}}lccl}
\toprule
\textbf{Metric} & \textbf{Unaided Condition} & \textbf{AI-Assisted} & \textbf{Difference} \\
 & \textbf{Mean $\pm$ s.d.} & \textbf{Mean $\pm$ s.d.} & \\
\midrule
\multicolumn{4}{c}{\textbf{Overall Performance (n=12)}} \\
\midrule
Accuracy (\%) & 16.5 $\pm$ 9.2 & 54.4 $\pm$ 11.9 & +37.9 \\
Response Time (s) & 45.0 $\pm$ 18.7 & 37.8 $\pm$ 15.1 & -7.2 \\
Confidence (1--5 scale) & 3.0 $\pm$ 0.7 & 4.0 $\pm$ 0.5 & +1.0 \\
\midrule
\multicolumn{4}{c}{\textbf{By Participant Expertise}} \\
\midrule
\multicolumn{4}{l}{\textit{Graduate Students (n=4)}} \\
Accuracy (\%) & 27.1 $\pm$ 7.3 & 65.7 $\pm$ 7.7 & +38.6 \\
Response Time (s) & 45.9 $\pm$ 17.3 & 39.1 $\pm$ 16.8 & -6.8 \\
Confidence (1--5 scale) & 3.4 $\pm$ 0.5 & 4.0 $\pm$ 0.4 & +0.6 \\
\addlinespace
\multicolumn{4}{l}{\textit{Lay Users (n=8)}} \\
Accuracy (\%) & 11.2 $\pm$ 4.0 & 48.7 $\pm$ 9.3 & +37.5 \\
Response Time (s) & 44.5 $\pm$ 20.5 & 37.1 $\pm$ 15.3 & -7.4 \\
Confidence (1--5 scale) & 2.8 $\pm$ 0.7 & 4.0 $\pm$ 0.5 & +1.2 \\
\bottomrule
\end{tabular*}
\end{table}

We evaluated whether retrieval-based assistance in {\systemName} improves human interpretation of OBS in a controlled, within-subject study. Twelve participants---4 graduate students with training in OBS palaeography, and 8 lay participants (university-educated native Chinese speakers with no prior exposure to OBS or ancient script studies)---completed 50 trials each drawn from a benchmark test set. Each participant performed both unaided and AI-assisted trials. In the assisted condition, the interface displayed the top-10 candidate modern characters, deduplicated by modern character label and ranked by cosine similarity, together with their synthesized OBS-style variants. We recorded top-1 accuracy (exact match to the canonical modern character label), mean response time per trial (s), and per-trial confidence on a 1--5 Likert scale (1 = very uncertain; 5 = very certain). Unless noted, values are mean~$\pm$~s.d. across participants. Results are shown in Table~\ref{tab:human_evaluation_new}.

Assistance improved overall accuracy (measured as exact match to the canonical modern character label in the benchmark; this evaluates structural matching, not full decipherment) by 37.9 pp, from 16.5\% to 54.4\% (a 229\% relative increase), reduced mean response time by 7.2~s (45.0~s to 37.8~s; 16\% faster), and increased confidence by 1.0 point (3.0 to 4.0). Among graduate students, accuracy improved from 27.1\% to 65.7\% (+38.6~pp; 142\% relative increase), with response times decreasing by 6.8~s (45.9~s to 39.1~s) and confidence increasing by 0.6 points (3.4 to 4.0). Among lay participants, accuracy improved from 11.2\% to 48.7\% (+37.5~pp; 335\% relative increase), with response times decreasing by 7.4~s (44.5~s to 37.1~s) and confidence increasing by 1.2 points (2.8 to 4.0). Absolute gains were comparable across groups, with larger relative improvements for lay participants; given the small specialist sample (n = 4), between-group contrasts should be interpreted cautiously.

These outcomes are consistent with the intended mechanism of {\systemName}. By synthesizing plausible OBS-style variants from modern characters and retrieving candidates, the interface externalizes evidence and narrows the search space. Participants compare stroke structure and component layout against ranked, high-probability hypotheses rather than scanning the full label space, which aligns with the observed pattern of higher accuracy, shorter decision times, and increased confidence.

In sum, retrieval-based, interpretable assistance augmented human performance without displacing expert judgment: participants were more accurate, faster, and more confident when assisted, and the benefits were evident across expertise levels while preserving transparency and scholarly oversight.

\section{Discussion}
\label{sec:disc}

Deciphering ancient scripts such as OBS requires methods that learn from few labels and remain reliable on damaged material. We cast OBS decipherment as retrieval against a generative, human‑auditable dictionary rather than as direct classification. {\systemName} synthesizes OBS‑style variants from modern Chinese characters and retrieves over this large dictionary, attaining high recall in zero‑shot settings (no task‑specific training on the target corpus) while preserving visible evidence for each suggestion. On HUST‑OBS and EVOBC, performance scales from strong Top‑10 shortlists to near‑exhaustive Top‑100 recovery and remains robust on degraded inputs. In a controlled user study, the system improved experts’ accuracy, speed and confidence without obscuring their judgment.

A central finding is the gap between our results and published claims for pipelines that map OBS images directly to modern characters via OCR. In our benchmarking, both released baselines and careful re‑implementations underperformed the original reports at test time, despite repeated trials and matched preprocessing and data splits (Supplementary Information). These discrepancies persisted under controlled conditions, indicating limited robustness and generalization beyond the models’ native training regimes.

Two mechanisms explain this fragility. First, the mapping direction creates an ill‑posed problem because the class spaces are highly imbalanced: OBS spans roughly 4{,}600 classes, whereas the full Unicode-encoded CJK ideograph inventory exceeds 87{,}000 entries~\citep{zhang2025megahan97k}---a collection spanning characters from diverse historical periods and regional traditions, far exceeding the few thousand in active modern use. Learning from the smaller to the much larger space means many modern targets have no reliable OBS predecessor. The model is therefore forced to guess among many indistinguishable options, yielding information loss and unstable decision boundaries when test characters are unseen. Second, these pipelines end with a closed‑set classifier---i.e., a model that must choose from a fixed list of known labels. Such classifiers cannot accommodate characters absent from their training data, including rare or undigitized forms; when applied to newly excavated or stylistically unconventional glyphs, this constraint further restricts generalization and compounds the class‑space mismatch.

Our approach avoids these failure modes by inverting the mapping and replacing closed‑set classification with retrieval. We synthesize OBS‑style variants from modern characters using a two‑stage generator---FAD followed by SR guided by the IDS---and index them in a shared embedding space for image retrieval. Inverting the mapping acts as a dimensionality‑reduction step, condensing modern character semantics into a smaller OBS‑style hypothesis set and mitigating incomplete‑coverage issues. Retrieval confers open‑set behavior: new items become searchable upon encoding without retraining, and Top‑$N$ shortlists remain stable as the corpus grows.

Several important limitations frame appropriate use and scope of interpretation. First, regarding the scope of the structural matching approach: OBS decipherment is a multifaceted process that involves determining the pronunciation, meaning, and linguistic properties of the words written by ancient graphs~\citep{takashima2000towards}. Our system addresses one specific aspect---identifying modern-style characters with structurally similar graphic forms---which generates hypotheses that must then be evaluated through broader philological, phonological, and contextual analysis. We acknowledge that many undeciphered OBS graphs may not be structurally isomorphic to any modern character: some write words that have disappeared from the language, while others have undergone such extensive graphic evolution that no modern counterpart shares recognizable components. Our current method is most directly applicable to cases where structural continuity exists between OBS and modern forms. Future extensions incorporating partial component matching, spatial rearrangement tolerance, and phonetic inference could broaden applicability to the more challenging cases that dominate the remaining undeciphered corpus.

Second, dictionary coverage and diversity depend on priors---fonts, stroke grammars, IDS quality and variant counts---and may bias retrieval toward common evolutionary trajectories; rare allographs could be underrepresented. Iterative dictionary refinement improves recall but risks confirmation drift; our procedure refines only dictionary entries, preserves test inputs and re-embeds each iteration, and future safeguards could include encoder freezing, validation-based early stopping on withheld characters and expert audits. Top-1 remains moderate even after refinement (47.5\%), reinforcing that the right unit of assistance is an interpretable shortlist. Heavy blur and noise impose a legibility threshold by suppressing discriminative morphology; specialized deblurring or denoising and multi-view may help.

More broadly, the combination of mapping inversion, component‑aware synthesis and image retrieval offers a general strategy for open‑vocabulary inference in historical scripts. It sidesteps the closed‑set constraints and class‑space mismatch that limit traditional pipelines \citep{guan2024deciphering}, scales with growing corpora without retraining, and provides audit trails via explicit ``evolutionary pathways''. These properties are pertinent not only to OBS decipherment but also to historical OCR and related settings where labels are sparse, forms are variable and discovery is ongoing.

% Appendices removed for arXiv preprint

\bibliography{sn-bibliography}

@article{marchant2025ai,
  title={How {AI} is unlocking ancient texts—and could rewrite history},
  author={Marchant, Jo},
  journal={Nature},
  volume={637},
  number={8044},
  pages={14--17},
  year={2025}
}

@article{assael2022restoring,
  title={Restoring and attributing ancient texts using deep neural networks},
  author={Assael, Yannis and Sommerschield, Thea and Shillingford, Brendan and Bordbar, Mahyar and Pavlopoulos, John and Chatzipanagiotou, Marita and Androutsopoulos, Ion and Prag, Jonathan and De Freitas, Nando},
  journal={Nature},
  volume={603},
  number={7900},
  pages={280--283},
  year={2022},
  publisher={Nature Publishing Group UK London}
}

@article{assael2025contextualising,
  title={Contextualising ancient texts with generative neural networks},
  author={Assael, Yannis and Sommerschield, Thea and Cooley, Alison and Pavlopoulos, John and Shillingford, Brendan and Herms, Bailey and Suresh, Priyanka and Maynard, Benjamin and Grayston, Justin and Wulgaert, Robbe and others},
  journal={Nature},
  volume={638},
  pages={1187--1194},
  year={2025},
  doi={10.1038/s41586-024-08481-0},
  publisher={Nature Publishing Group UK London}
}

@article{boltz1986early,
  title={Early {Chinese} writing},
  author={Boltz, William G},
  journal={World Archaeology},
  volume={17},
  number={3},
  pages={420--436},
  year={1986},
  publisher={Taylor \& Francis}
}

@article{keightley1979shang,
  title={The {Shang} state as seen in the oracle-bone inscriptions},
  author={Keightley, David N},
  journal={Early China},
  volume={5},
  pages={25--34},
  year={1979},
  publisher={Cambridge University Press}
}

@article{guo2015building,
  title={Building hierarchical representations for oracle character and sketch recognition},
  author={Guo, Jun and Wang, Changhu and Roman-Rangel, Edgar and Chao, Hongyang and Rui, Yong},
  journal={IEEE Transactions on Image Processing},
  volume={25},
  number={1},
  pages={104--118},
  year={2015},
  publisher={IEEE}
}

@article{liu2020oracle,
  title={Oracle bone inscriptions recognition based on deep convolutional neural network ({CNN})},
  author={Liu, Mengting and Liu, Guoying and Liu, Yongge and Jiao, Qingju},
  journal={Journal of image and graphics},
  volume={8},
  number={4},
  pages={114--119},
  year={2020}
}

@article{li2025comprehensive,
  title={A comprehensive survey of oracle character recognition: Challenges, datasets, methodology, and beyond},
  author={Li, Jing and Chi, Xueke and Wang, Qiufeng and Huang, Kaizhu and Wang, Da-Han and Liu, Yongge and Liu, Cheng-Lin},
  journal={Pattern Recognition},
  pages={111824},
  year={2025},
  publisher={Elsevier}
}

@book{bazerman2009handbook,
  title        = {Handbook of Research on Writing: History, Society, School, Individual, Text},
  author       = {Bazerman, Charles},
  year         = {2009},
  publisher    = {Routledge},
  address      = {New York; London}
}

@article{zhang2025megahan97k,
  title={{MegaHan97K}: A large-scale dataset for mega-category {Chinese} character recognition with over 97{K} categories},
  author={Zhang, Yuyi and Shi, Yongxin and Zhang, Peirong and Zhao, Yixin and Yang, Zhenhua and Jin, Lianwen},
  journal={Pattern Recognition},
  pages={111757},
  year={2025},
  publisher={Elsevier}
}

@article{zhen1995astronomy,
  title={Astronomy on oracle bone inscriptions},
  author={Zhen-Tao, Xu and Stephenson, FR and Yao-Tiao, Jiang},
  journal={Quarterly Journal of the Royal Astronomical Society, Vol. 36, p. 397},
  volume={36},
  pages={397},
  year={1995}
}

@article{takashima2000towards,
  title={Towards a more rigorous methodology of deciphering oracle-bone inscriptions},
  author={Takashima, Ken’ichi},
  journal={T'oung Pao},
  volume={86},
  number={Fasc. 4/5},
  pages={363--399},
  year={2000},
  publisher={JSTOR}
}

@phdthesis{chen2009compound,
  title={Compound ideograph: a contested category in studies of the {Chinese} writing system},
  author={Chen, Zhiqun},
  year={2009},
  school={Monash University}
}

@inproceedings{liu2022convnet,
  title={A convnet for the 2020s},
  author={Liu, Zhuang and Mao, Hanzi and Wu, Chao-Yuan and Feichtenhofer, Christoph and Darrell, Trevor and Xie, Saining},
  booktitle={Proceedings of the IEEE/CVF conference on computer vision and pattern recognition},
  pages={11976--11986},
  year={2022}
}

@article{ho2020denoising,
  title={Denoising diffusion probabilistic models},
  author={Ho, Jonathan and Jain, Ajay and Abbeel, Pieter},
  journal={Advances in neural information processing systems},
  volume={33},
  pages={6840--6851},
  year={2020}
}

@inproceedings{ronneberger2015u,
  title={U-net: Convolutional networks for biomedical image segmentation},
  author={Ronneberger, Olaf and Fischer, Philipp and Brox, Thomas},
  booktitle={International Conference on Medical image computing and computer-assisted intervention},
  pages={234--241},
  year={2015},
  organization={Springer}
}

@book{wang1999reading,
  title={Reading {Chinese} script: A cognitive analysis},
  author={Wang, Jian and Chen, Hsuan-Chih and Radach, Ralph and Inhoff, Albrecht},
  year={1999},
  publisher={Psychology Press}
}

@article{reichle2018models,
  title={Models of {Chinese} reading: Review and analysis},
  author={Reichle, Erik D and Yu, Lili},
  journal={Cognitive Science},
  volume={42},
  pages={1154--1165},
  year={2018},
  publisher={Wiley Online Library}
}

@inbook{Boltz+2021+845+864,
url = {https://doi.org/10.1515/9783110753301-041},
title = {Textual Criticism and Early Chinese Manuscripts},
booktitle = {Exploring Written Artefacts},
booktitle = {Objects, Methods, and Concepts},
author = {William G. Boltz},
editor = {Jörg B. Quenzer},
publisher = {De Gruyter},
address = {Berlin, Boston},
pages = {845--864},
doi = {doi:10.1515/9783110753301-041},
isbn = {9783110753301},
year = {2021},
}

@article{wang2024open,
  title={An open dataset for {Oracle Bone Script} recognition and decipherment},
  author={Wang, Pengjie and Zhang, Kaile and Wang, Xinyu and Han, Shengwei and Liu, Yongge and Wan, Jinpeng and Guan, Haisu and Kuang, Zhebin and Jin, Lianwen and Bai, Xiang and others},
  journal={arXiv preprint arXiv:2401.15365},
  year={2024}
}

@article{guan2024open,
  title={An open dataset for the evolution of oracle bone characters: {EVOBC}},
  author={Guan, Haisu and Wan, Jinpeng and Liu, Yuliang and Wang, Pengjie and Zhang, Kaile and Kuang, Zhebin and Wang, Xinyu and Bai, Xiang and Jin, Lianwen},
  journal={arXiv preprint arXiv:2401.12467},
  year={2024}
}

@article{dhariwal2021diffusion,
  title={Diffusion models beat {GANs} on image synthesis},
  author={Dhariwal, Prafulla and Nichol, Alexander},
  journal={Advances in neural information processing systems},
  volume={34},
  pages={8780--8794},
  year={2021}
}

@inproceedings{zhu2017unpaired,
  title={Unpaired image-to-image translation using cycle-consistent adversarial networks},
  author={Zhu, Jun-Yan and Park, Taesung and Isola, Phillip and Efros, Alexei A},
  booktitle={Proceedings of the IEEE international conference on computer vision},
  pages={2223--2232},
  year={2017}
}

@article{saharia2022image,
  title={Image super-resolution via iterative refinement},
  author={Saharia, Chitwan and Ho, Jonathan and Chan, William and Salimans, Tim and Fleet, David J and Norouzi, Mohammad},
  journal={IEEE transactions on pattern analysis and machine intelligence},
  volume={45},
  number={4},
  pages={4713--4726},
  year={2022},
  publisher={IEEE}
}

@inproceedings{li2023bbdm,
  title={Bbdm: Image-to-image translation with brownian bridge diffusion models},
  author={Li, Bo and Xue, Kaitao and Liu, Bin and Lai, Yu-Kun},
  booktitle={Proceedings of the IEEE/CVF conference on computer vision and pattern Recognition},
  pages={1952--1961},
  year={2023}
}

@inproceedings{guan2024deciphering,
  title={Deciphering {Oracle Bone} Language with Diffusion Models},
  author={Guan, Haisu and Yang, Huanxin and Wang, Xinyu and Han, Shengwei and Liu, Yongge and Jin, Lianwen and Bai, Xiang and Liu, Yuliang},
  booktitle={Proceedings of the 62nd Annual Meeting of the Association for Computational Linguistics (Volume 1: Long Papers)},
  pages={15554--15567},
  year={2024}
}

@article{Wu2025,
author = { Wu, Shengya },
title = {Supplementary Interpretation of the Oracle Bone Script Character ``qiang'' [{Ji\v{a}g\v{u}w\'{e}n} ``{qi\={a}ng}'' z\`{i} b\v{u}sh\`{i} / \begin{CJK}{UTF8}{gbsn}甲骨文"羌"字补释\end{CJK}]},
journal = {Jianghan Archaeology [Ji\={a}ngh\`{a}n K\v{a}og\v{u} / \begin{CJK}{UTF8}{gbsn}江汉考古\end{CJK}]},
volume = {},
number = {01},
pages = {174-178},
year = {2025},
issn = {1001-0327},
}

@article{NiuLiu2023,
author = { Niu, Jianmin and Liu, Zhengwen},
title = {Textual Criticism on {Chinese} Oracle of ``{t\={a}o}'' and ``\begin{CJK}{UTF8}{gbsn}妻\end{CJK}'' [{Ji\v{a}g\v{u}w\'{e}n} ``{t\={a}o}'' y\v{u} ``{q\={\i}}'' k\v{a}obi\`{a}n]},
journal = {Journal of Anyang Institute of Technology [\={A}ny\'{a}ng G\={o}ngxu\'{e}yu\`{a}n Xu\'{e}b\`{a}o / \begin{CJK}{UTF8}{gbsn}安阳工学院学报\end{CJK}]},
volume = {22},
number = {01},
pages = {119-121},
year = {2023},
issn = {1673-2928},
doi = {10.19329/j.cnki.1673-2928.2023.01.029}
}

@article{Yuan2022,
author = { Yuan, Lunqiang },
title = {Supplementary Explanations to {L\"{u}} in Oracle Bone Inscriptions [{Ji\v{a}g\v{u}w\'{e}n} ``{l\v{u}}'' z\`{i} b\v{u}shu\={o} / \begin{CJK}{UTF8}{gbsn}甲骨文"履"字补说\end{CJK}]},
journal = {Unearthed Literature [Ch\={u}t\v{u} W\'{e}nxi\`{a}n / \begin{CJK}{UTF8}{gbsn}出土文献\end{CJK}]},
volume = {},
number = {02},
pages = {43-50+154},
year = {2022},
issn = {2096-7365},
}

@article{Hou2024,
author = { Hou, Naifeng },
title = {Textual Criticism on Oracle ``{ch\'{u}}'' [{Ji\v{a}g\v{u}w\'{e}n} ``{ch\'{u}}'' z\`{i} k\v{a}o / \begin{CJK}{UTF8}{gbsn}甲骨文"雏"字考\end{CJK}]},
journal = {Bulletin of Oracle Bone Inscriptions and Yin-Shang History [Ji\v{a}g\v{u}w\'{e}n y\v{u} Y\={\i}nSh\={a}ng Sh\v{i} Y\'{a}nji\={u} / \begin{CJK}{UTF8}{gbsn}甲骨文与殷商史研究\end{CJK}]},
number = {00},
pages = {348-360},
year = {2024},
}

@article{diao2025oracle,
  title={Oracle bone inscription image restoration via glyph extraction},
  author={Diao, Xiaolei and Shi, Daqian and Cao, Wei and Wang, Ting and Qi, Ruihua and Li, Chuntao and Xu, Hao},
  journal={npj Heritage Science},
  volume={13},
  number={1},
  pages={321},
  year={2025},
  publisher={Springer International Publishing Cham}
}

@article{wang2025chinese,
  title={{Chinese} inscription restoration based on artificial intelligent models},
  author={Wang, Zhen and Li, Yujun and Li, Honglei},
  journal={npj Heritage Science},
  volume={13},
  number={1},
  pages={326},
  year={2025},
  publisher={Springer International Publishing Cham}
}

@inproceedings{yang2024fontdiffuser,
  title={Fontdiffuser: One-shot font generation via denoising diffusion with multi-scale content aggregation and style contrastive learning},
  author={Yang, Zhenhua and Peng, Dezhi and Kong, Yuxin and Zhang, Yuyi and Yao, Cong and Jin, Lianwen},
  booktitle={Proceedings of the AAAI conference on artificial intelligence},
  volume={38},
  number={7},
  pages={6603--6611},
  year={2024}
}

@article{heusel2017gans,
  title={{GANs} trained by a two time-scale update rule converge to a local {Nash} equilibrium},
  author={Heusel, Martin and Ramsauer, Hubert and Unterthiner, Thomas and Nessler, Bernhard and Hochreiter, Sepp},
  journal={Advances in neural information processing systems},
  volume={30},
  year={2017}
}

@article{li2011recognition,
  title={Recognition of inscriptions on bones or tortoise shells based on graph isomorphism},
  author={Li, Qingsheng and Yang, Yuxing and Wang, Aimin},
  journal={Jisuanji Gongcheng yu Yingyong(Computer Engineering and Applications)},
  volume={47},
  number={8},
  pages={112--114},
  year={2011},
  publisher={North China Computing Technology Institute,| a No. 26, P. O. Box 619| c~…}
}

@inproceedings{meng2017recognition,
  title={Recognition of Oracle Bone Inscriptions by Extracting Line Features on Image Processing.},
  author={Meng, Lin},
  booktitle={ICPRAM},
  pages={606--611},
  year={2017}
}

@article{gan2023characters,
  title={Characters as graphs: Interpretable handwritten {Chinese} character recognition via {Pyramid Graph Transformer}},
  author={Gan, Ji and Chen, Yuyan and Hu, Bo and Leng, Jiaxu and Wang, Weiqiang and Gao, Xinbo},
  journal={Pattern Recognition},
  volume={137},
  pages={109317},
  year={2023},
  publisher={Elsevier}
}
\end{document}